\documentclass[a4paper]{jpconf}
\usepackage{graphicx}
\usepackage[square,sort&compress, numbers]{natbib}
\usepackage{citesort}

\begin{document}
\title{Recent bounds on graviton mass using galaxy clusters}

\author{Shantanu Desai}

\address{Dept. of Physics, IIT Hyderabad, Kandi Telangana-502285, India}

\ead{shantanud@iith.ac.in}

\author{Sajal Gupta}
\address{Department of Physical Sciences, IISER-Kolkata, Mohanpur, West Bengal-741246, India}
\ead{sg15ms084@iiserkol.ac.in}

\begin{abstract}
Although galaxy clusters have proved to be wonderful laboratories for testing a plethora of  modified gravity theories and other exotic alternatives to $\Lambda$CDM, until a few years ago, there was only one paper (from   1974), which obtained a limit on graviton mass (of $\mathcal{O}(10^{-29})$ eV) with clusters.  To rectify this, in the last few years multiple works have obtained different bounds on graviton mass using single cluster data as well as stacking galaxy catalogs. We review these recent limits on graviton mass using galaxy clusters obtained using disparate methods. 
 \end{abstract}

\section{Introduction}
Even though General relativity agrees with all observational tests~\cite{Will,Ishak,Ferreira},  a large number of  alternate gravity theories have been proposed after the equations of General Relativity were written down, more than a century ago. The main motivation is  to resolve cosmological problems of dark energy, dark matter, baryogenesis, inflation as well as data-driven problems such as Hubble constant tension or $\sigma_8$ tension~\cite{Verde}. Aside from cosmological considerations, another motivation for pursuing modified gravity theories is to address certain conceptual problems such as arrow of time, big-Bang singularity, quantization of gravity etc.~\cite{Desai18}.

Modified theories of gravity, where graviton is endowed is with a non-zero mass have been in vogue, ever since they were first proposed by Pauli and Fierz in 1939~\cite{Goldhaber10}. Such theories can resolve problems with dark energy, dark matter, inflation, quantization of gravity etc. In the 1970's, conceptual problems with these theories were raised such as vDVZ discontinuity and Bouleware-Deser ghosts~\cite{Goldhaber10}. In the last decade, there has been a breakthrough in these problems, leading to  a resurgence of interest in massive gravity theories~\cite{derham11,derham14}.

A recent review on graviton mass bounds using different observational probes can be found in Refs.~\citep{Derham16,PDG}. There are three different ways to constrain the graviton mass~\cite{Derham16}. The first method involves looking for a weakening of the gravitational force due to a Yukawa  potential. The second constraint comes from  looking for fifth force interactions, which arise in massive gravity models.  The third type of limit  comes from the propagation of gravitational waves,  either  due to  modified dispersion relations or from difference in arrival times between gravitational waves and photons.

The first bound on graviton mass (from the first method) using galaxy clusters was obtained by Goldhaber and Nieto in 1974~\cite{Goldhaber74}. Galaxy clusters are the most massive gravitationally collapsed objects in the universe~\cite{Voit}, and have  proved to be invaluable for testing modified theories of gravity and other cosmological parameters at the interface of fundamental Physics, such as non-gaussianity and neutrino mass. On the observational front, a large number of new galaxy clusters up to very high redshifts have been discovered in the past two decaudes  courtesy dedicated surveys in optical, X-ray, and microwave (through S-Z effect) from Stage-II and Stage-III dark energy experiments. This will continue into the next decade because of multiple   stage IV dark energy experiments such as LSST, Euclid, WFIRST etc. are about to start taking data.

Despite  the huge progress on the observational front for galaxy clusters, there was no updated limit on graviton mass from galaxy clusters for the last four decades after Ref.~\cite{Goldhaber74}. This situation was soon rectified starting in 2018, when new limits  on graviton mass with galaxy clusters using latest observational data were presented to supersede the old limit from 1974~\cite{Desai18,Rana,Gupta,Gupta19}.  

In this  paper we provide a succinct review of these updated limits on graviton mass using clusters, along with the first ever limit. We do not discuss limits on graviton mass from Weak Lensing, solar system, gravitational waves or  using the black hole in the center of our galaxy. These can be found in recent reviews~\cite{Derham16,PDG} or citations therein.

\section{Goldhaber and Nieto limit}

Goldhaber and Nieto~\cite{Goldhaber74} used the fact that galaxy clusters are bound and the largest clusters known then  (from the Holmberg catalog) had sizes of around 580 kpc. According to Bertrand's theorem, only Newtonian gravity gives rise to closed bound orbits. Using the fact that any violations from Newtonian gravity must be valid only at distances greater than the maximum separation between clusters (building on similar ideas earlier on by Hare~\cite{Hare}), a bound on graviton mass was obtained by positing $e^{-1} \leq \exp(-\mu_g r)$, where  $\mu_g$ is the inverse of the reduced Compton wavelength. From this, the estimated limit is  $m_g<1.1\times10^{-29}$ eV, or in terms of graviton Compton wavelength is given by $\lambda_g>10^{20}$ km. For more than four decades, this was the only bound on graviton mass from clusters and has been  widely cited in literature (although not even once in cosmology/galaxy clusters literature). Occasionally, concerns have been raised about this limit because of uncertainties related to dark matter distribution~\cite{Will97}. However, this criticism is invalid, since this limit  does not make any assumption about the  dark matter or any other mass distribution with clusters.

\section{Recent cluster-based bounds}
We now review recent works on obtained a limit on graviton mass using clusters starting from 2018.
\subsection{Limit from Abell 1689}
The Abell 1689 cluster is one of the largest 
and most massive galaxy cluster located at a redshift of 0.18. In the past decade, a whole slew of X-ray, lensing and SZ observations have provided very precise dynamical mass models for this cluster.  This cluster has been a poster child to test a large number of alternate gravity theories and alternatives to $\Lambda$CDM~\cite{Nieu}. From the  dynamical mass models for the gas, dark matter and density profile derived in Ref.~\cite{Nieu}, the acceleration was computed as a function of distance from the center of the cluster for both Newtonian and Yukawa gravity. A $\chi^2$ residual was constructed  between the two accelerations and including the reconstructed errors in acceleration. The 90\% c.l. limit on graviton mass was obtained from 
$\Delta \chi^2 < 2.71$. This limit corresponds to $m_g<1.37\times 10^{-29}$ eV or $\lambda_g>9.1 \times 10^{19}$ km. More details related to this bound can be found in Ref.~\cite{Desai18}.

\subsection{Limits from stacked cluster catalogs}
Rana et al~\cite{Rana} considered a  catalog of 182  galaxy clusters detected by the ACT-SZ survey as well as a  catalog of 50 clusters from the LoCuSS Weak lensing survey. For each cluster, they calculated the $\chi^2$ residuals from the difference in Yukawa and Newtonian acceleration for a given  cluster mass, followed by summing over all clusters in the survey. The error in the acceleration was obtained from a quadrature sum of the errors in mass and the errors in the Hubble parameter. The upper limit on graviton mass was obtained from this $\chi^2$ to obtain upper bounds at 68\%, 90\%,95\% and 99\% confidence intervals.
The most stringent upper limit obtained from this analysis was from 
the weak lensing catalog, corresponding to   $m_g <7.85 \times 10^{30}$ eV or $\lambda_g> 1.579 \times 10^{20}$ km~\cite{Rana}. This same method was applied to the SPT-SZ, Planck-SZ, SDSS-redMaPPer catalog consisting of about 500, 900,  and 26000  clusters respectively. The best limit from the SDSS-redMapper catalog, corresponding to a $m_g<1.27\times10^{-30}$ eV or $\lambda_g>9.76\times 10^{20}$ km~\cite{Gupta}. Using this same method, the sensitivity of the Euclid cluster catalog woudl lead a graviton mass bound of about $10^{-31}$ eV.

\subsection{Limits from Chandra X-ray catalog}

Mostly recently, a catalog of 12 relaxed clusters from the Chandra X-ray sample, for which detailed temperature and density profiles were obtained by Vikhlinin et al~\cite{Vikhlinin06} was used to bound the graviton mass~\cite{Gupta19}. From the equation of hydrostatic equilibrium in a Yukawa potential, the total Yukawa hydrostatic mass was then estimated. The $\chi^2$ residual was constructed from the   deviations between  this Yukawa mass and Newtonian mass and including the errors in the Newtonian mass. Using this method, the best limit was obtained for Abell 2390 corresponding to $m_g<3.46 \times 10^{-29}$ eV or $\lambda_g>3.58\times 10^{19}$ km~\cite{Gupta19}. This same method should be applicable to the galaxy clusters from the recently launched eRosita X-ray telescope.

\section{Conclusions}
In this short proceedings, we recapitulate all limits which have been obtained on the graviton mass using galaxy clusters. All these limits assume that in the weak field limit a non-zero graviton mass gives rise to a Yukawa potential. The first ever limit was obtained using the fact that any deviations from Newtonian gravity kick in at length scales greater than 580 kpc~\cite{Goldhaber74}.  Then in 2018, a limit was obtained from the Abell 1689 galaxy cluster using the difference in Yukawa and Newtonian accelerations. This was followed by limits obtained using stacked galaxy cluster catalogs from optical and SZ surveys using similar technique. Finally, a  catalog of 12 clusters observed with the Chandra X-ray telescope  was used to obtain the most recent limits.

A tabular summary of all these results can be found in Table~\ref{tab:1}. The best limit comes from the stacked catalog of SDSS redMaPPer clusters, corresponding to a limit of  $m_g<1.27\times10^{-30}$ eV or $\lambda_g>9.76\times 10^{20}$ km~\cite{Gupta}.

\begin{table}[h]
\centering
\caption{\label{tab:1} A summary of various bounds on graviton mass using galaxy clusters. Note that for the Chandra X-ray catalog we only quote the cluster with the most stringent upper limit. All limits are at 90\% c.l. except the first one for which no confidence limit was provided.}

\begin{tabular}{|l|c|c|}
\br
Reference & Cluster/Catalog & Limit on graviton mass (eV)\\
\mr
Goldhaber and Nieto~\cite{Goldhaber74}&Holmberg & $1.1 \times 10^{-29}$ \\
Desai~\cite{Desai18} & Abell  & $1.37 \times 10^{-29}$  \\
Rana et al~\cite{Rana} & LocuSS (WL)  & $7.849 \times 10^{-30}$  \\
Rana et al~\cite{Rana} & ACT  (SZ) & $1.05 \times 10^{-29}$ \\
Gupta and Desai~\cite{Gupta} & SDSS-RedMaPPer & $1.27 \times 10^{-30}$ \\
Gupta and Desai~\cite{Gupta} & SPT (SZ) & $4.73\times 10^{-30}$  \\
Gupta and Desai~\cite{Gupta} & Planck (SZ)  & $3 \times 10^{-30}$  \\
Gupta and Desai~\cite{Gupta19} & Abell 2390  & $3.46 \times 10^{-29}$  \\
\br
\end{tabular}
\end{table}

\bibliography{main}

\end{document}